# Numerical Simulation of Wavy-Flap Airfoil Performance at Low Reynolds Number: Insights from Lift and Drag Coefficient Analysis

Mohammad Amin Esabat, MME[1] *, Saeed Jaamei, Ph.D.[2], Fatemeh Asadi, MME[3], Ahmad Reza Kohansal, Ph.D.[4], and Hassan Abyn, Ph.D.[5]

[1] Researcher, Marine Department, Persian Gulf University, Bushehr, Iran, 7516913817; email: Esabat@mehr.pgu.ac.ir (Corresponding author)
[2] Associate Professor, Marine Department, Persian Gulf University, Bushehr, Iran, 7516913817 email: jaameisa@pgu.ac.ir
[4] Associate Professor, Marine Department, Persian Gulf University, Bushehr, Iran, 7516913817 email: kohansal@pgu.ac.ir
[5] Associate Professor, Marine Department, Persian Gulf University, Bushehr, Iran, 7516913817 email: abynhassan@pgu.ac.ir
[3] Ph.D. Candidate, Ocean Engineering Department, Texas A&M University, Galveston, USA, 77554 email:fatemeasaditb@tamu.edu

**Abstract**. This research examines the aerodynamic performance of wavy (corrugated) airfoils, focusing specifically on analyzing the impact of two angles of attack: the airfoil's angle of attack and the tail's angle of attack (β). Simulations were conducted using the W1011 airfoil at a Reynolds number of 200,000, considering attack angles of 0°, 2°, 5°, and 8° for the airfoil, and 0°, 10°, 20°, 30°, and 40° for the tail. The simulation outcomes were validated against experimental data from Williamson's laboratory.The findings indicate a notable increase in the lift coefficient for the wavy airfoils, especially at larger flap angles. Specifically, at a β of 40°, the lift coefficient of the wavy airfoil at a 0° angle of attack was nearly three times greater than in the other scenarios. In contrast, the drag coefficient also increased, but to a lesser extent, which suggests enhanced aerodynamic efficiency. Furthermore, the lift-to-drag ratio was significantly higher for the wavy airfoils, particularly at lower angles of attack. However, at increased angles of attack, the performance enhancements diminished, with a slight reduction in the lift-to-drag ratio due to the development of turbulence and low-pressure zones on the airfoil's surface. Overall, the study concludes that wavy airfoils, especially with higher tail angles, offer considerable aerodynamic benefits, especially in conditions of low angle attack, improving lift and fuel efficiency for use in both aviation and marine applications.

*List of Abbreviations*

- $\alpha$     Airfoil angle of attack
- $\beta$     Airfoil tail angle of attack
- $cl$     Lift coefficient
- $cd$     Drag coefficient
- $s$     Airfoil studied in laboratory results (W1011)
- $w$     Airfoil studied (modified)

# 1. INTRODUCTION

The aerodynamic performance of wave-shaped airfoils has attracted significant attention, particularly in marine applications where noise reduction and aerodynamic efficiency are crucial. This reduction not only enhances operational performance but also helps meet stringent noise regulations in maritime environments. Numerous studies on airfoils with wavy leading edges have demonstrated that significant amplitude and minimal wavelength can substantially moderate noise levels (Chen 2022). For instance, Rarata (2023) compared smooth and wavy airfoils, observing that while modifications to the suction side did not affect the acoustic feedback loop mechanism, disturbances on the pressure side reduced trailing-edge noise. Further research by Zhang et al. (2020) highlighted that undulating surfaces and protrusions on airfoils can mitigate aerodynamic stalls, extending the operational lifespan of these surfaces.

Bio-inspired wavy leading edges have also shown promise in reducing trailing-edge bluntness noise, achieving reductions as high as 35 dB (Xing et al. 2023a). Spanwise wavy surfaces have been found to enhance aerodynamics, reducing trailing-edge noise by up to 17.7 dB in some cases, while also lowering pressure fluctuations and drag (Smith and Klettner 2022). In a related study, Xing et al. (2023b) further investigated the impact of bio-inspired wavy leading edges, concluding that airfoils with greater amplitude and smaller wavelength were more effective at noise reduction, achieving a decrease of up to 33.9 dB in sound pressure levels. Ferreira et al. (2022) also explored the effects of wave combinations on cylinder aerodynamics, noting a reduction in drag, vortex formation, and separation bubbles, inspired by biological structures.

Chen et al. (2024) expanded on the use of wavy leading edges, examining their effect on noise reduction in turbulent flow. Their study compared three bionic treatments-wavy, porous, and combined revealing that although these modifications increased mean drag, they successfully reduced both broadband and tonal noise. Additional studies found that grooved and wavy leading-edge structures effectively reduced mid-frequency noise, with simulations showing that such structures enhanced decorrelation and reduced pulsation pressure, contributing to better noise control (Sun et al. 2023).

Although wavy leading edges have proven effective at reducing noise, they can adversely affect aerodynamic performance by increasing drag and decreasing lift. However, these modifications have been shown to reduce lift fluctuations, aerodynamic instability, and tonal noise. Studies have demonstrated that sinusoidal trailing-edge serrations can reduce broadband noise across various frequencies by disrupting vortex shedding and diminishing surface pressure fluctuations (Singh and Narayanan 2023). Inspired by owl wings and humpback whale fins, leading-edge serrations have been shown to reduce airfoil-turbulence interaction noise, with the largest wave amplitude yielding a 2.2 dB reduction in mid-frequency noise (Yu et al. 2024).

Additionally, research on supercritical airfoils with undulating leading edges has suggested that controlling wave amplitude is crucial for improving aerodynamic performance and reducing shock buffet (Degregori and Kim 2021). Du et al. (2024) investigated the effect of wavy leading edges (WLEs) on airflow patterns, revealing that these surfaces could reduce separation at high angles of attack, thereby improving the flow attachment and overall aerodynamic stability. Furthermore, studies of the NACA 4412 airfoil under undulating conditions have highlighted the potential risks of large amplitude waves, which can exacerbate aerodynamic instability and separation at certain Reynolds numbers (Liu et al. 2021).

Hu and Ma (2020)'s study on the NACA 4412 airfoil's aerodynamic properties under undulating ground conditions revealed regular variations in coefficients, except when the angle of attack was 0°. The study recommends using the middle angle of attack, with undulating terrain causing most changes in aerodynamic coefficients. This information can be useful for wing-in-ground craft design. A study comparing NACA 4412 airfoil's aerodynamics subjected to wavy water surfaces and stiff ground found significant differences influenced by Reynolds and Froude numbers and distinct flow processes, providing guidance for safe wing-in-ground craft flight design (Hu et al. 2021). The interaction between airflow and sea waves significantly influences ground-effect vehicle flight stability, and circulation control during relative upturn can enhance safety and stability (Hao et al 2022).

Zverkov and Kryukov (2021) reduced airfoil-turbulence interaction noise using edge teeth inspired by owl wings and humpback whale fins, achieving a 2.2 dB reduction in mid-frequency noise. A new technique using leading-edge protuberances (LEPs) improves wind turbine blade performance by managing flow separation, enhancing aerodynamic performance and reducing flow separation, thereby enhancing airfoil design efficiency (Yi-Nan et al. 2021). Experiments on a NACA 0012 airfoil showed that as the angle of attack increases, laminar boundary layer instability noise changes from broadband hump to tonal noise, which then becomes a broadband hump with an increasing Reynolds number (Abedi and Zakeri 2024, Güzey et al. 2024) . Gao et al. (2022) conducted a numerical simulation on the NACA 0012 airfoil, examining wavy roughness features near the leading edge. The study found that roughness affects pressure, skin friction coefficients, and separation behavior, but does not significantly alter the transition's start. The bionic airfoil design, inspired by humpback whales, reduces blade vibration, dynamic stall, drag coefficient, lift coefficient. Enhancing the waviness ratio enhances dynamic hysteresis effect (Wu and Liu 2021). Xin et al. (2022)'s numerical simulations explored pitching airfoil behavior in flat and undulating ground, revealing kinematic characteristics and flow dynamics, influencing wake vortices formation and force peaks. Airfoils with undulating leading edges improve aerodynamic performance and minimize noise (Wang et al. 2024). This study involved numerical simulations at 50,000 Reynolds number (Wang et al. 2022). Zhao et al. (2022) conducted a large eddy simulation on the impact of undulating leading edges on cavitation management in a modified NACA634-021 hydrofoil, finding no significant decrease in cavity volume. AI technology also enhances flow management options, but understanding flow behavior is crucial. A framework for optimizing the geometry of airfoils using machine learning with the goals of increasing resistance to flutter and decreasing drag has been developed (Jung ang Gu 2024).

Despite the growing body of work on wavy airfoils, there remains a need for more comprehensive studies that specifically address the aerodynamic performance of these surfaces at low Reynolds numbers. The present study aims to bridge this gap by examining the effects of wavy airfoils on lift, drag, and efficiency at a Reynolds number of 200,000. The focus will be on analyzing the influence of wave characteristics—such as undulation heights and pattern—on aerodynamic performance. Additionally, the study will explore the interaction between the airfoil and the ground effect, utilizing the Navier-Stokes equations and large eddy simulations (LES) to model turbulent flow conditions. Preliminary results indicate that the flap angle has a more significant impact on the lift coefficient than the airfoil's angle of attack, with modifications to the wavy surface showing a notable increase in lift, albeit with a corresponding rise in drag. These findings provide valuable insights for optimizing wavy airfoils for marine vessel applications at low Reynolds numbers.

## 2. METHODS

### 2.1 Governing Equations and Numerical Methods for Near-Surface Airfoil Interactions

The aerodynamic performance of airfoils in proximity to a surface, such as the ground or water, is influenced by complex flow dynamics that require careful mathematical modeling and numerical analysis. The accuracy and reliability of the results depend heavily on the governing equations used to describe the flow as well as the numerical methods implemented to solve them. In this study, the primary set of equations governing the flow are the Navier-Stokes equations, which describe the motion of fluid substances. These equations, in conjunction with specific turbulence models and numerical techniques, are used to simulate and understand the effects of near-surface interactions on the aerodynamic behavior of the wavy airfoils at low Reynolds numbers.

*2.1.1 Governing Equations: Navier-Stokes Equations*

The flow of fluids with high viscosity is governed by the Navier-Stokes equations, a set of partial differential equations that describe the motion of incompressible and compressible fluids. These equations are fundamental in fluid mechanics, as they model the conservation of mass, momentum, and energy in a flow field. The Navier-Stokes equations are derived from the application of Newton's second law of motion to fluid motion, with the assumption that the stress in the fluid is proportional to the rate of change of velocity (i.e., the velocity gradient). In contrast to the Euler equations, which are used to model inviscid flow (i.e., flows with negligible viscosity), the Navier-Stokes equations account for the effects of viscosity, making them suitable for a wide range of fluid flow problems, including turbulent and boundary-layer flows.

Mathematically, the Navier-Stokes equations can be expressed as a system of equations that consist of the continuity equation (representing mass conservation), the momentum equation (which describes the forces and accelerations in the flow), and the energy equation (which governs thermal effects). In analyzing the fluid flow around airfoils, the governing equations are essential to describing the dynamics of the system. The continuity equation (mass conservation) ensures that mass is conserved throughout the flow field (Eq. 1). For incompressible fluids, it is expressed as:

$$\frac{\partial \rho}{\partial t} + \nabla \cdot (\rho u) = 0 \tag{1}$$

where u is the velocity vector. This equation indicates that the net mass flow into any region of the fluid must be balanced by the flow out of that region, implying that mass does not accumulate or deplete.

The momentum equation, also known as the Navier-Stokes equation, governs the conservation of momentum in the fluid, considering the effects of pressure gradients, viscous forces, and external forces. It is given by (Eq. 2):

$$\frac{\partial(\rho u)}{\partial t} + \nabla \cdot (\rho u \otimes u) = -\nabla p + \nabla \cdot (\mu \nabla u) + f \tag{2}$$

where $p$ represents the pressure, $\mu$ is the dynamic viscosity, and $f$ includes external forces (e.g., gravity or electromagnetic forces). The term $\nabla^2 u$ captures the viscous diffusion of momentum, while the nonlinear term $u \otimes u$ represents the advection, or convective transport of momentum.

Finally, the energy equation describes the conservation of energy, accounting for both heat conduction and convection within the flow. It is written as (eq. 3):

$$\frac{\partial(\rho E)}{\partial t} + \nabla \cdot (\rho u H) = \nabla \cdot (\mu \nabla u) + \nabla \cdot (\kappa \nabla T) + \dot{q} \tag{3}$$

where $T$ is the temperature, $\kappa$ is the thermal conductivity, and $q$ represents any heat source terms. This equation balances the rate of change of thermal energy within the fluid with the diffusion of heat due to conduction and the advective transport of thermal energy due to the flow.

These equations, when solved together, provide a comprehensive description of the fluid's behavior, enabling the analysis of various flow characteristics around the airfoil, especially in proximity to the ground or other surfaces where effects like ground proximity or wall interactions can significantly alter the flow dynamics (Kocić et al. 2020).

In the context of airfoils, the Navier-Stokes equations are particularly important when the airfoil operates near a stationary surface, such as the ground or water. This proximity introduces the ground effect, which reduces aerodynamic drag and enhances lift. Ground effect occurs when the downward flow (downwash) behind the wing is restricted by the surface, inhibiting tip vortex formation and reducing induced drag. This phenomenon significantly alters the flow behavior, especially near the surface, and is crucial for understanding the performance of airfoils used in ground-effect vehicles (GEVs) or marine vessels. For example, during takeoff, a fixed-wing aircraft may float just above the runway, benefiting from ground effect to accelerate until it reaches the climb speed. The effect is most pronounced when the airfoil is at or below half its wingspan above the surface, resulting in improved lift-to-drag ratios and enhanced aerodynamic performance

*2.1.2 Numerical Methods: Finite Volume Approach*

The Finite Volume Method (FVM) is employed to solve the Navier-Stokes equations and model the interactions between the airfoil and the ground. The method is particularly advantageous for computational fluid dynamics (CFD) simulations due to its inherent ability to conserve mass, momentum, and energy within control volumes, making it ideal for complex geometries and unstructured grids.

In FVM, the computational domain is discretized into small, finite volumes. The governing equations are then solved within these control volumes, where the fluxes of conserved quantities (e.g., mass, momentum, energy) are computed across the surfaces of each control volume. The key strength of FVM lies in its ability to ensure that conservation laws are strictly adhered to at the discrete level. A time-stepping algorithm is used to advance the solution forward in time, allowing for the simulation of dynamic flow behavior.

For a scalar quantity $\phi$ (such as temperature, concentration, or pressure), the finite volume discretization of the governing equations is expressed in the following general form of (Eq. 4):

$$\int_V \frac{\partial \phi}{\partial t}\, dV + \oint_S \mathbf{F} \cdot \mathbf{n}\, dA = 0 \qquad (4)$$

where *V* is the volume of the control volume, *S* represents the surface of the control volume, *F* is the flux of the scalar quantity ϕ, and *n* is the unit vector normal to the surface *S*. The sum of the fluxes across all surfaces of the control volume is considered, ensuring the correct transport and conservation of the scalar quantity across the domain.

This approach is well-suited for fluid dynamics simulations because it allows for accurate representation of complex flow fields, including the effects of turbulence, surface interactions, and other physical phenomena critical to understanding the behavior of airfoils in proximity to surfaces (Ghalambaz et al. 2022).

*2.1.3 Computational Considerations: Grid Resolution and Turbulence Modeling*

The accuracy of a simulation largely depends on its grid resolution and turbulence modeling. To precisely capture the boundary layer effects near the airfoil and the ground, a "sufficiently fine grid" is employed. In numerical simulations, particularly in computational fluid dynamics (CFD) for airfoil analysis, determining a sufficiently fine grid is crucial for accurately modeling physical phenomena. This involves considering several factors: the analysis goals, critical physical regions needing high precision, and the turbulence models used. The grid must be fine enough to capture details in boundary layers and flow separation while balancing computational cost. Convergence analysis helps verify result stability across different grid densities, and industry standards provide guidance. Ultimately, we refine the grid step by step to strike the optimal balance between precision and resources.

In this study, Large Eddy Simulation (LES) is chosen for simulating the flow around an airfoil using the Finite Volume Method (FVM) because it balances accuracy and computational feasibility in modeling turbulent flows. LES gives a more accurate picture of complicated flow phenomena like separation and vortex shedding because it can resolve larger eddies. This is very important for understanding how unsteady flow dynamics work. It is particularly adept at predicting time-dependent flow behaviors, essential for analyzing the dynamic response of airfoils. LES offers improved accuracy over RANS by modeling only the smallest scales, delivering detailed insights into flow structures and turbulence mechanisms. While more computationally demanding, LES's precision and comprehensive flow information are invaluable for precise aerodynamic analyses and design optimizations. Eq. (5) expresses the filtered Navier-Stokes equations in the case of large eddy simulation (LES).

$$\frac{\partial \tilde{u}}{\partial t} + (\tilde{u} \cdot \nabla)\tilde{u} = -\frac{1}{\rho}\nabla \tilde{p} + \nu \nabla^2 \tilde{u} - \nabla \cdot \tau + f \tag{5}$$

where $\tilde{u}$ represents the filtered velocity field, $\tilde{p}$ is the filtered pressure, $\rho$ is the fluid density, $\nu$ is the kinematic viscosity, $\tau$ denotes the subgrid-scale stress tensor, and $f$ represents external forces (Mohamad et al. 2020).

### 2.2 Geometry and Meshing

The geometry of the airfoils has been designed and adjusted according to the specifications outlined in Figure 1. Three distinct airfoil models have been created, each featuring different undulation heights s, allowing for a comparative analysis of the effects of varying wave amplitudes on airfoil performance. The models explore how changes in undulation heights influence fluid flow dynamics, particularly turbulence and boundary layer behavior.

Small undulation heights s are expected to induce subtle local effects and may slightly increase turbulence, while larger undulation heights s are more likely to generate significant flow changes and vortices, which could lead to more pronounced aerodynamic effects. This study aims to assess the sensitivity of the flow to these geometric changes by simulating fluid dynamics using the Navier-Stokes equations. By investigating the impact of small versus large undulations on airfoil performance, the research seeks to determine whether minor adjustments can yield results similar to those of more substantial changes, ultimately aiding in the optimization of aerodynamic design.

The airfoil models are categorized based on the ratio of the tail chord length to the total airfoil chord length. Specifically, the models are grouped into three categories with ratios of 0.1, 0.25, and 0.5. The largest undulation heights corresponds to the model with a tail chord-to-total airfoil chord ratio of 0.5, yielding a maximum undulation heights of 4.5 cm. The smallest undulation heights is 9 mm. Additionally, distances from the tail to the edge of the airfoil are represented as percentages of the total airfoil chord: 100%, 97%, 92.5%, and 85%. These values reflect the spanwise distribution of the undulation heights along the airfoil.

The horizontal axis in the geometry plot corresponds to the spanwise ratio relative to the airfoil's chord length, while the vertical axis represents the distance from the trailing edge to the leading edge of the airfoil, again expressed as a percentage of the chord. This geometric setup allows for a clear understanding of how variations in undulation heights and placement influence the overall aerodynamic performance of the airfoil.

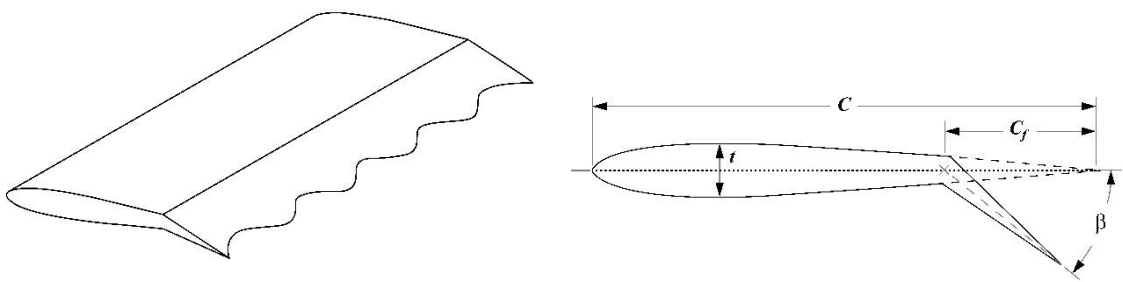

**Fig. 1** Isometric and Side view of the airfoil geometry

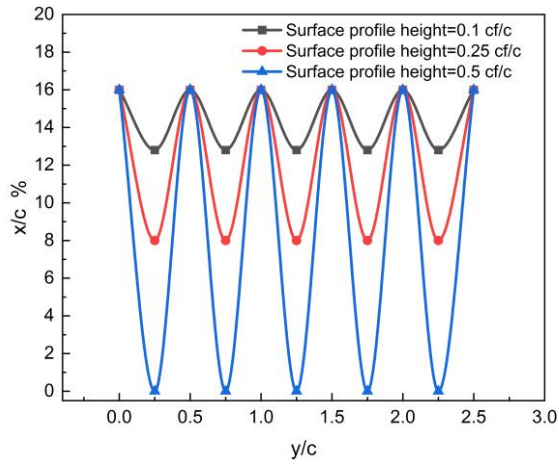

**Fig. 2** Geometry of the modified airfoil (W1011) showing undulation heights.

**Table 1** Mesh independence analysis showing the number of mesh elements for each simulation.

| Simuluation number | Mesh |
|---|---|
| 1 | 452000 |
| 2 | 890000 |
| 3 | 2800000 |
| 4 | 3700000 |

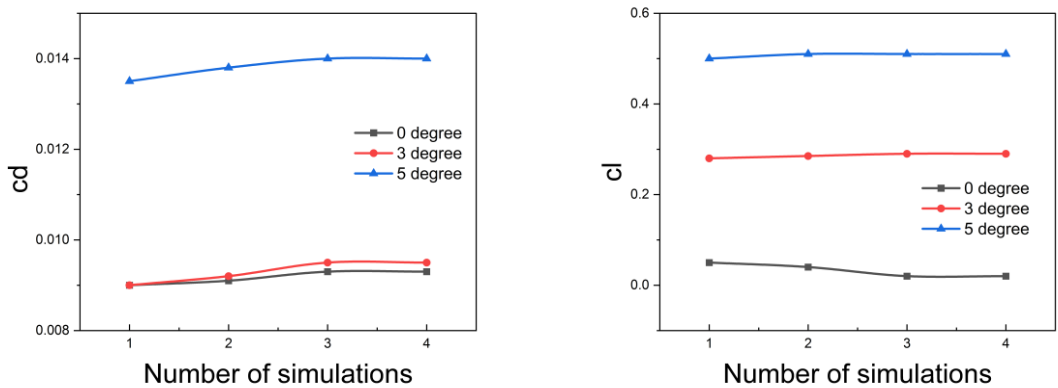

**Fig. 3** Drag and lift coefficients diagram of mesh-independent simulations

To ensure the results are solely a consequence of the applied physics and boundary conditions, rather than the mesh resolution, mesh independence studies were conducted in the computational fluid dynamics (CFD) simulations. Mesh independence is defined as the point at which further refinement of the mesh does not significantly alter the simulation outcomes. As shown in Table 1, the mesh was simulated four times, and after evaluating the results, a mesh of 3,700,000 cells was selected as the optimal resolution. This mesh size was chosen because further increases in cell count did not significantly affect the velocity fields, pressure distribution, or aerodynamic coefficients such as lift and drag. The mesh independence study ensures that the simulation results are reliable and not influenced by variations in mesh density.

**Table 2** Grid mesh and airfoil size parameters for different mesh refinement settings.

|   | Mesh Refinement | Wake mesh | Surface mesh |
|---|---|---|---|
| 1 | 0.002 | 0.01 | 0.004 |
| 2 | 0.002 | 0.01 | 0.04 |

Table 2 shows the details of the selected mesh sizes used in the simulation, specifically the wake and surface mesh refinements. These mesh sizes are crucial for accurately capturing the flow behavior around the airfoil. Table 3, on the other hand, presents the general geometry dimensions of the simulated airfoil, providing essential information on its size and shape for context in the simulation study.

**Table 3**: Dimensions of the computational domain (tank) used in the simulations.

| Axis | Length (m) |
|---|---|
| Along x-axis | 10.5 |
| Along y-axis | 4.0 |
| Along z-axis | 1.5 |

To validate the accuracy of the results, the W1011 airfoil was simulated at a Reynolds number of 200,000 with angles of attack of 0°, 2°, and 5°. Additionally, simulations for the airfoil flap were conducted across a range of angles of attack, specifically 0°, 10°, 20°, 30°, and 40°. The simulation results were then compared with experimental data from Williamson's laboratory (Williamson 2012).

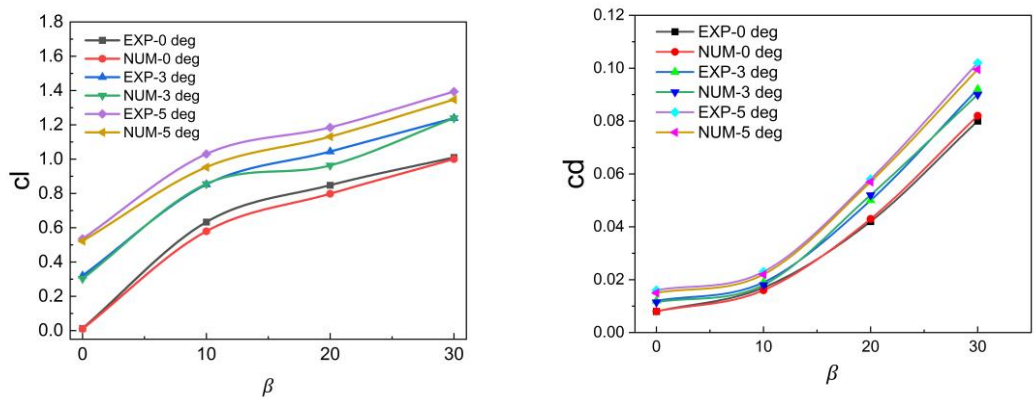

**Fig. 4** Lift and Drag coefficient comparison between simulation results and experimental data from Williamson's laboratory (Williamson 2012).

Figure 5 provides a 3D visualization of the simulation tank alongside the geometry of the wavy W1011 airfoil. This representation illustrates the setup of the simulation, highlighting the modified airfoil's design with undulating features, as well as the spatial configuration of the computational domain used for the fluid dynamics analysis.

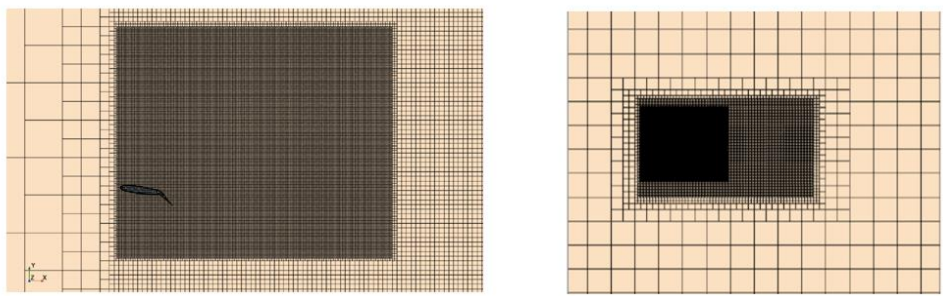

**Fig. 5** 3D view of the simulation tank and the geometry of the wavy W1011 airfoil.

# 3. RESULTS

It is important to clarify that in this study, two distinct angles of attack are considered based on Williamson's tests. One angle of attack is defined by the airfoil itself, while the second angle is determined by the tail of the airfoil. In the following diagrams, the term "$\alpha$" refers to the angle of attack of the airfoil, whereas the term "$\beta$" refers to the angle of attack of the tail relative to the airfoil's direction, as illustrated in Figure 1.

In our study, the chord length ($c_f$) is set to 0.3 times the airfoil chord, and the airfoil used is the W1011. The angles of attack examined include 0°, 3°, 5°, and 8° for the airfoil, and 0°, 10°, 20°, 30°, and 40° for the tail.

Figure 6 presents the lift coefficient diagram, comparing the simulation results with Williamson's experimental data. The diagram covers airfoil angles of attack of 0°, 3°, 5°, and 8°, with the horizontal axis indicating the angle of attack of the W1011 airfoil's tail ($\beta$). The percentages in the diagram represent the undulation heights relative to the tail chord, where 10% corresponds to a undulation heights of 9 mm (i.e., 10% of the tail's chord length). The undulation heights s for the other cases are 0.9 cm, 2.25 cm, and 4.5 cm.

In the diagram, a significant increase in the lift coefficient for the wavy airfoils, compared to the natural (flat) airfoil, is evident. For instance, at a $\beta$ of 40°, the airfoil at a 0° attack angle shows a lift coefficient approximately three times greater than that of the natural airfoil. At a $\beta$ of 10°, the same increase is observed, reaching a 4.2-fold growth at the same attack angle. This effect becomes even more pronounced at higher tail angles, with a greater increase in lift.

Interestingly, the influence of $\beta$ on the lift coefficient is more substantial than the airfoil's own angle of attack. Although the positive impact of $\beta$ is clear, the graphs show that as the airfoil's angle of attack increases particularly at angles of 3°, 5°, and 8° the growth ratio tends to stabilize. For example, at a 40° $\beta$ angle, at 8° of attack, the lift coefficient increases by 2.6 times for a undulation heights of 50%, while at 3° and 0° attack angles, the growth ratios are 2.7 and 2.5 times, respectively, compared to the natural airfoil.

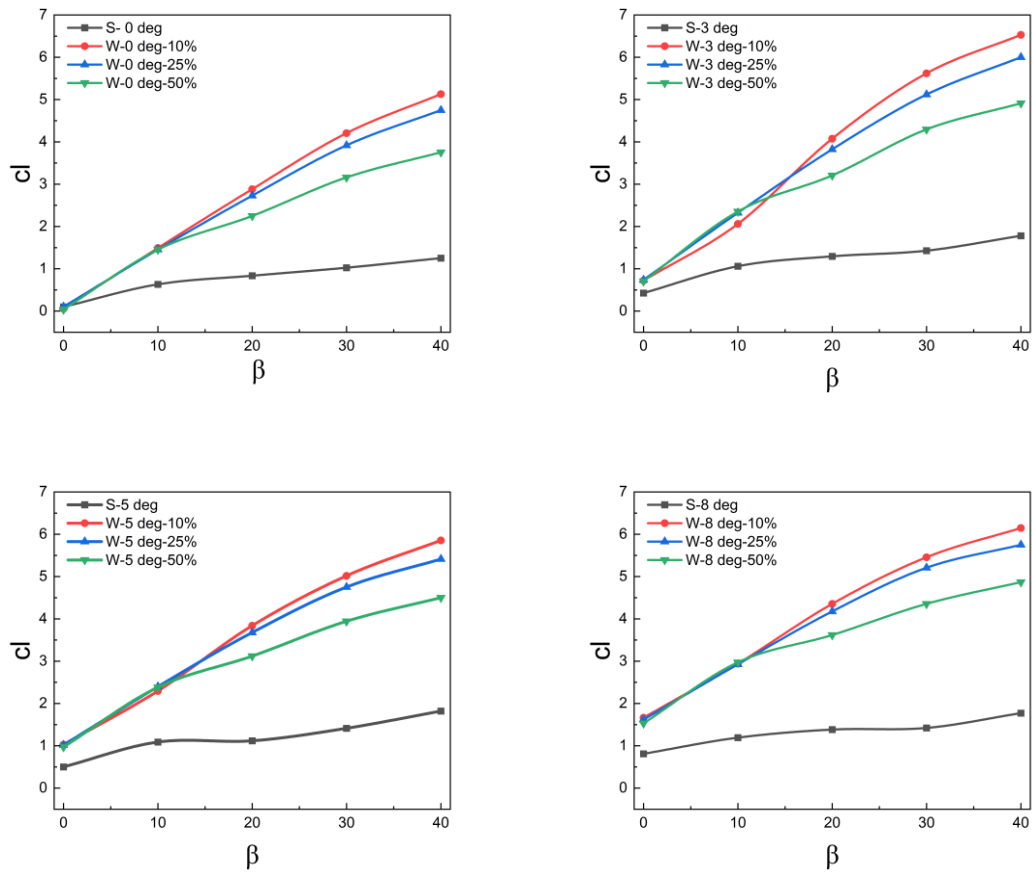

**Fig. 6** Lift coefficient diagram of wavy airfoils and W1011

In fluid dynamics, the drag coefficient is a dimensionless quantity used to measure an object's resistance to motion through a fluid, such as air or water. A lower drag coefficient indicates that the object experiences less drag. This coefficient is related to the object's surface area and is composed of skin friction and form drag, the primary contributors to fluid dynamic drag. For lifting bodies such as airfoils or hydrofoils, the drag coefficient also accounts for lift-induced drag. Additionally, for more complex structures like airplanes, interference drag is considered.

In contrast, the lift coefficient is a dimensionless number that relates the lift produced by a body to the surrounding fluid's density, the body's velocity, and a reference area. A lifting body, such as a fixed-wing aircraft or any body with foils, generates lift. The lift coefficient can be calculated using parameters such as Mach number, Reynolds number, and the angle of attack to the flow. For two-dimensional foils, the section lift coefficient represents dynamic lift properties, with the foil chord serving as the reference area.

While the drag coefficient increases in a similar manner to the lift coefficient for wavy airfoils, its growth is not as pronounced. For example, as shown in Figure 7, with an angle of attack of 0 and 8 degrees and a $\beta$ of 20 degrees, the drag coefficient increases by 1.078 and 3.14 times, respectively, at a undulation heights of 10%. This indicates a significant influence of the angle of attack on the drag coefficient. However, while the angle of attack strongly influences the drag coefficient, its impact on the lift coefficient is less significant.

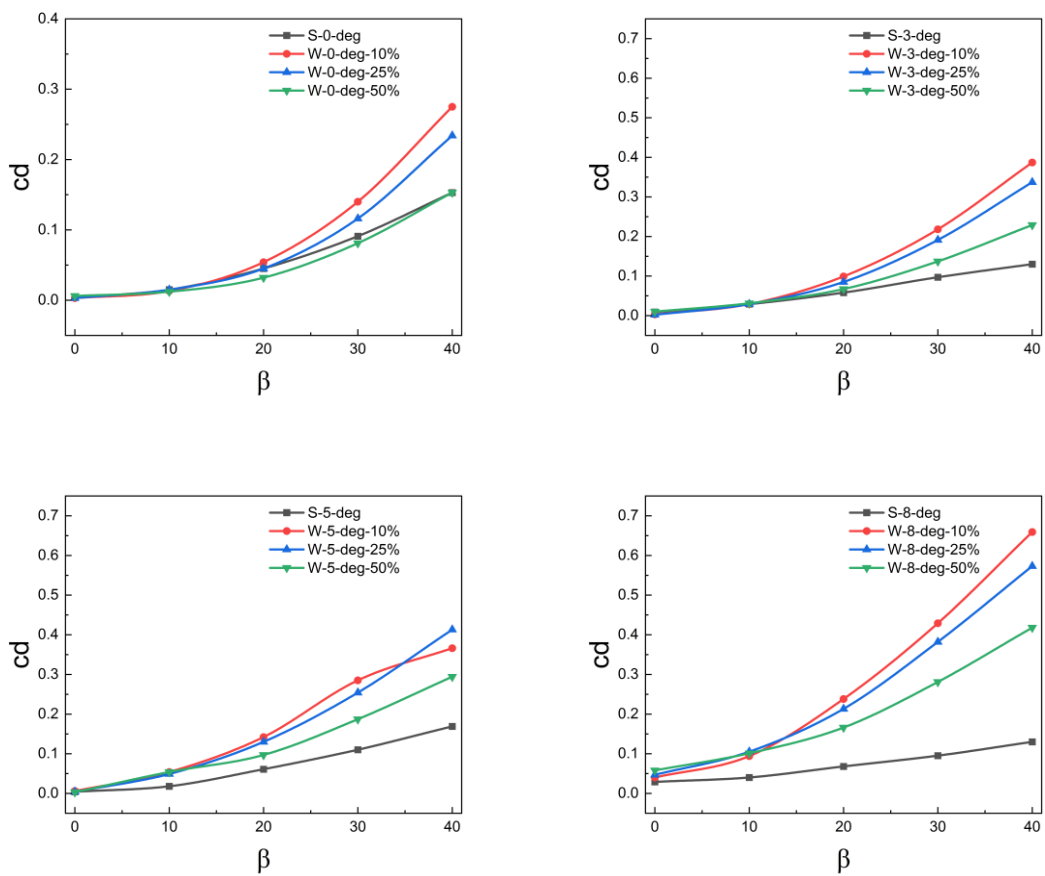

**Fig. 7** Drag coefficient diagram of wavy airfoils and W1011

Figure 8, which illustrates the lift-to-drag ratio, offers valuable insight into the performance of corrugated airfoils. The ratio increases at a slower rate as the angle of attack rises, indicating that the airfoils achieve better efficiency at lower angles of attack, particularly for lower β values. For instance, at an angle of attack of 10 degrees with zero degrees of β, the lift-to-drag ratio for the main airfoil is 40.5. In comparison, the wavy simulation models show a lift-to-drag ratio ranging from 92 to 100, reflecting a 2.2 to 2.5 times increase. At an angle of attack of 8° with β values of 40°, 30°, and 20°, the increase in the lift-to-drag ratio becomes negligible, suggesting that the influence of the flap angle diminishes at higher attack angles.

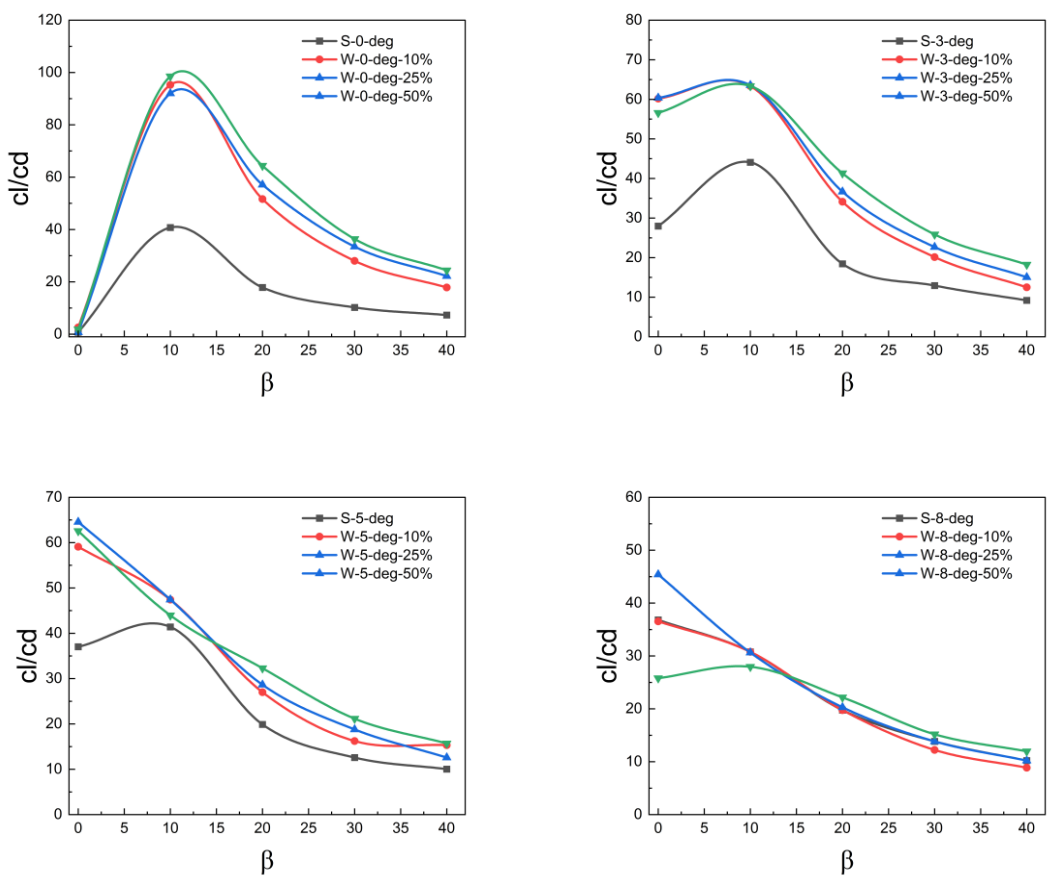

**Fig. 8** Lift-to-drag coefficient diagram of wavy airfoils and W1011

In certain cases, such as for the airfoil attack angle of zero degrees with a β of 0 degrees and a undulation heights of 50%, the lift-to-drag ratio was observed to be lower than that of the natural airfoil. However, this behavior is considered an exception in the performed simulations. In most cases, the wavy airfoils either show competitive performance or a significant increase in the lift-to-drag ratio compared to the original airfoil. Figure 9 demonstrates the same general trend for the ratio of lift-to-drag coefficients as the airfoil β increases. In this image, which consists of 5 graphs, the airfoil shows a wave at the airfoil angles of attack. In each graph, two lines are depicted corresponding to the wavy airfoils. The angles in each graph are related to the tail angles of attack, and the horizontal axis angles are related to the airfoil angles of attack. The percentages in each line also give information about the height of the wave at the tail end of the airfoils.

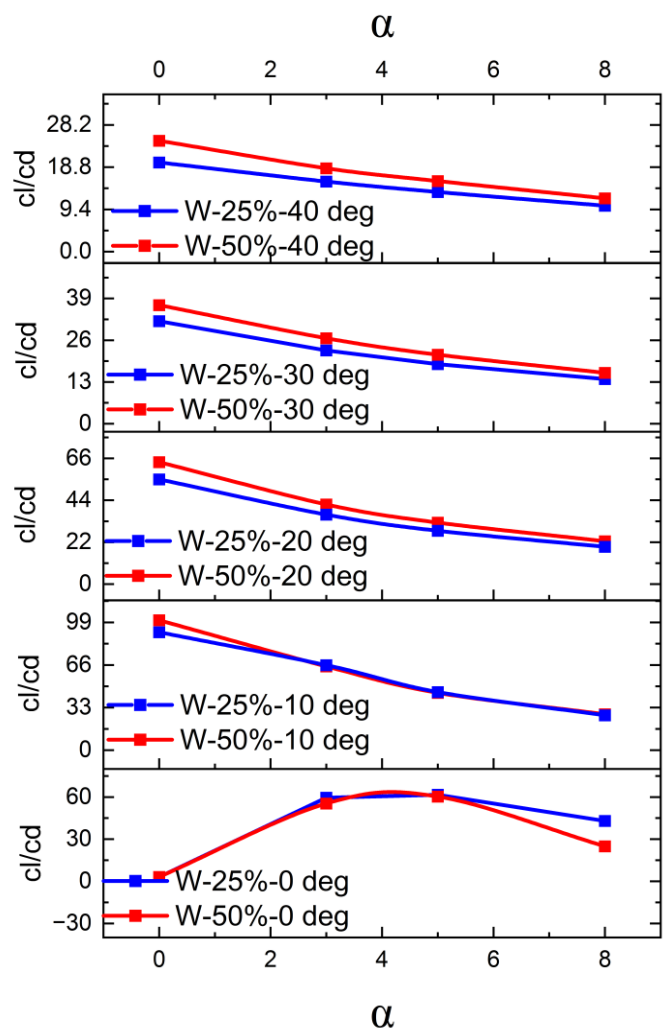

**Fig. 9** Lift-to-drag coefficient diagram of wavy airfoils and W1011. The percentages correspond to the undulation heights at the end of the airfoil (tail). The horizontal axis is the airfoil's $\alpha$, and each graph line is associated with $\beta$.

Figure 10 provides a clearer view of the simulation's contours by showcasing the corrugated W1011 airfoil in its tail section. It highlights three sections: A, B, and C, which correspond to the wave's trough, middle, and crest, respectively. For clarity, the terms bottom, middle, and top will be used to refer to Sections A, B, and C, respectively, moving forward in the text.

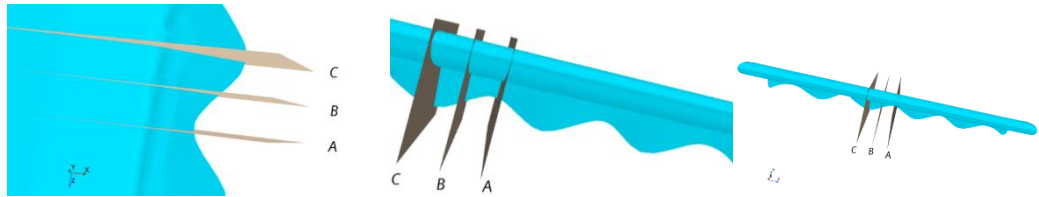

**Fig. 10** 3D view of the W1011 wavy airfoil, with Sections A, B, and C representing the trough, middle, and crest, respectively.

One of the primary reasons for the reduction in lift-to-drag ratio at higher angles of attack is the formation of low-pressure regions on the surface of the airfoils, coupled with turbulence flow. This phenomenon significantly contributes to the decrease in the ratio. Figures 11 and 12 illustrate this effect, showing the pressure distribution, streamline speed, vector movement, and vorticity around the wavy airfoil at two different β angles (0° and 40°), with the airfoil having an attack angle of 8°. At a β of 40°, the formation of vorticity is more pronounced on the surface of the airfoil, with a larger volume and higher amplitude of vorticity compared to the case at a 0° β. Additionally, Figure 13 presents a schematic of the 3D vorticity around the airfoil for both β angles (0° and 40°).

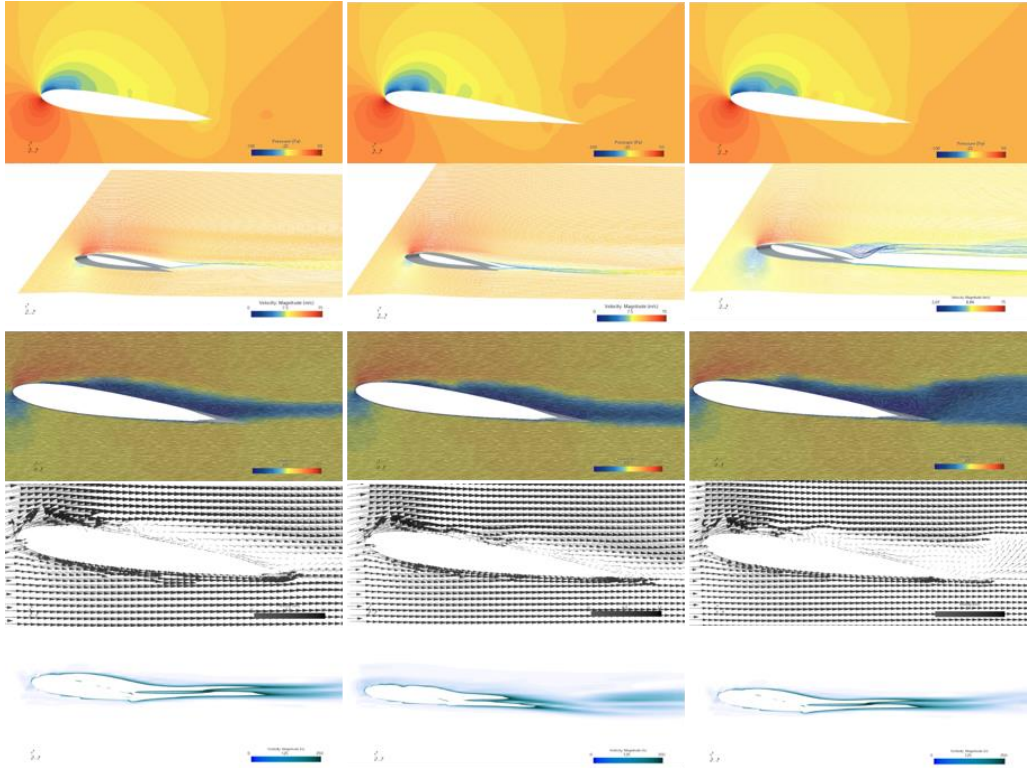

**Fig. 11** Contours in order of pressure, streamline velocity, scalar vector, velocity vector, and vorticity. The trough, middle, and crest sections are arranged from left to right, with α is 8° and $\beta$ is 0°.

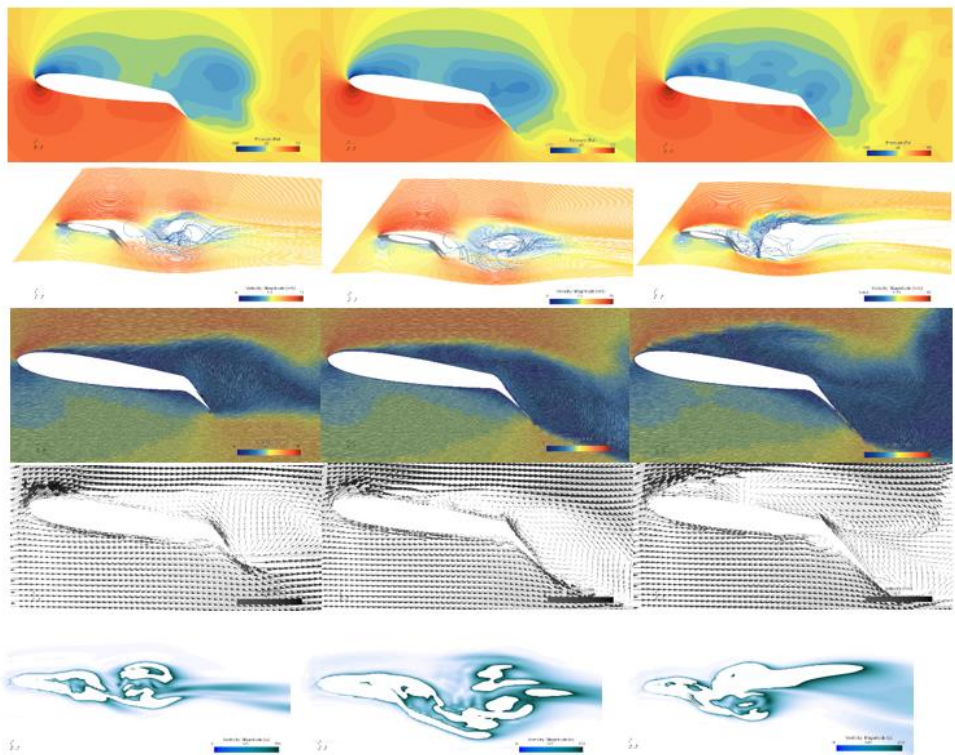

**Fig. 12** Contours in order of pressure, streamline velocity, scalar vector, velocity vector, and vorticity. The trough, middle, and crest sections are arranged from left to right, with α is 8° and $\beta$ is 40°.

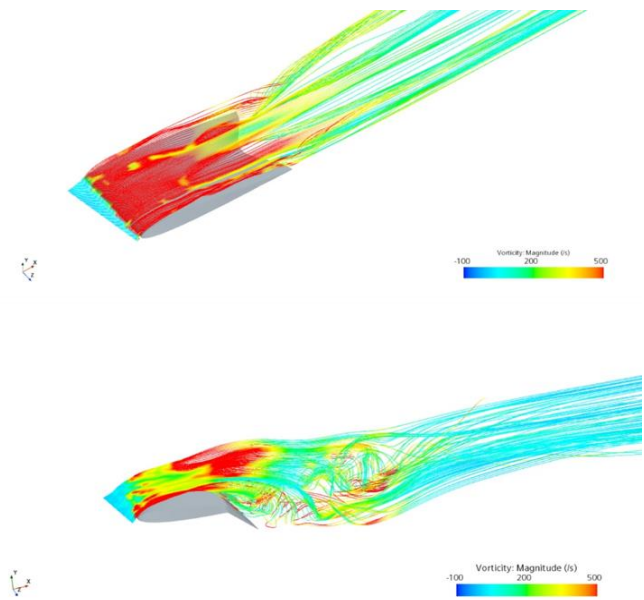

**Fig. 13** 3D vorticity streamline, showing a wavy airfoil with α is 8° and $\beta$ is 0°. The bottom image of the wavy airfoil α is 8° and $\beta$ is 40°

## 4. Conclusion

The present study aims to investigate the behavior of corrugated airfoils at two angles of attack: the angle of attack of the airfoil and the angle of attack of its tail. Williamson's tests were conducted to analyze the impact of these angles on the airfoil's lift coefficient. The results show that the tail's angle of attack significantly influences the lift coefficient.

The lift coefficient diagram indicates a significant increase in the lift coefficient of wavy airfoils compared to their natural state. The growth ratio increases considerably at higher tail angles. Similarly, the drag coefficient also increases in wavy airfoils, although not as significantly as the lift coefficient.

The ratio of lift-to-drag coefficients provides insightful information on the performance of corrugated airfoils. The lift-to-drag ratio shows less growth with an increase in the angle of attack, and it is more efficient at lower angles for lower $\beta$.

In conclusion, this study provides valuable insights into the behavior of corrugated airfoils and their impact on lift and drag coefficients. The findings indicate that the angle of attack of the tail has more influence on the lift coefficient than the angle of attack of the airfoil itself. These insights can help design more efficient airfoils for various applications.

Corrugated airfoils, with their unique aerodynamic properties, offer significant potential for engineering applications. Their ability to influence airflow due to their uneven surfaces can lead to improved aerodynamic characteristics such as reduced drag and increased lift. This is particularly important in simulations near the ground, where the effects of turbulent airflows and interactions with the ground surface are critical.

Continued research in this area is essential for several reasons:

1. Enhanced Stability and Control: A better understanding of how these airfoils interact with near-ground flows can improve the design of systems operating at low altitudes.

2. Multiple Industry Applications: From aerospace engineering to wind energy production, corrugated airfoils can drive innovation in turbine and energy generator designs, providing enhanced performance and efficiency.

**Ethical Approval**

This research did not contain any studies involving animal or human participants, nor did it take place on any private or protected areas. No specific permissions were required for corresponding locations.

**Competing Interests**

We have no competing interests.

**Funding**

There is no funding applicable to this article.

**Availability of Data and Materials**

# 5. REFERENCES

Abedi, M. and Zakeri, R., 2024. Experimental study of the aerodynamics of micro airfoils with wavy surface and the use of different patterns inspired by nature. *Journal of Technology in Aerospace Engineering*, pp.1-11.

Chen, W., Lei, H., Xing, Y., Wang, L., Zhou, T. and Qiao, W., 2024. On the airfoil leading-edge noise reduction using poro-wavy leading edges. *Physics of Fluids*, *36*(3).

Chen, W., Qiao, W., Duan, W. and Wei, Z., 2021. Experimental study of airfoil instability noise with wavy leading edges. *Applied Acoustics*, *172*, p.107671.

Degregori, E. and Kim, J.W., 2021. Mitigation of transonic shock buffet on a supercritical airfoil through wavy leading edges. *Physics of Fluids*, *33*(2).

Du, H., Jiang, H., Yang, Z., Xia, H., Chen, S. and Wu, J., 2024. Experimental Investigation of the Effect of Bio-Inspired Wavy Leading-Edges on Aerodynamic Performance and Flow Topologies of the Airfoil. *Aerospace*, *11*(3), p.194.

Ferreira, P.H., Moura, R.C. and de Araújo, T.B., 2022. Experimental Investigation of a Wavy Leading Edge Cylinder. In *AIAA AVIATION 2022 Forum* (p. 4041).

Gao, W., Samtaney, R. and Parsani, M., 2022. Direct numerical simulation of transitional flow past an airfoil with partially covered wavy roughness elements. *Physics of Fluids*, *34*(10).

Ghalambaz, M., Jin, H., Bagheri, A., Younis, O. and Wen, D., 2022. Convective flow and heat transfer of nano-encapsulated phase change material (NEPCM) dispersions along a vertical surface. *Facta Universitatis, Series: Mechanical Engineering*, *20*(3), pp.519-538.

Güzey, K., Aylı, U.E., Kocak, E. and Aradag, S., 2024. Investigation of aerodynamic and aeroacoustic behavior of bio-inspired airfoils with numerical and experimental methods. *Proceedings of the Institution of Mechanical Engineers, Part C: Journal of Mechanical Engineering Science*, *238*(5), pp.1265-1279.

Hao, L.I.U., Jianhong, S.U.N., Zhi, S.U.N., Yang, T.A.O., Dechen, W.A.N.G. and Guangyuan, L.I.U., 2022. Circulation control of airfoil aerodynamic force under ground effect of wavy wall. *Journal of Shanghai Jiaotong University*, *56*(8), p.1101.

Hu, H. and Ma, D., 2020. Airfoil aerodynamics in proximity to wavy ground for a wide range of angles of attack. *Applied Sciences*, *10*(19), p.6773.

Hu, H., Ma, D., Guo, Y. and Yang, M., 2021. Airfoil Aerodynamics in Proximity to Wavy Water Surface. *Journal of Aerospace Engineering*, *34*(2), p.04020119.

Jung, J. and Gu, G.X., 2024. Data-driven airfoil shape optimization framework for enhanced flutter performance. *Physics of Fluids*, *36*(10).

Kocić, M., Stamenković, Ž., Petrović, J. and Nikodijević, M., 2020. CONTROL OF MHD MICROPOLAR FLUID FLOW. *Facta Universitatis, Series: Automatic Control and Robotics*, *18*(3), pp.163-175.

Liu, X.A., Ma, D., Yang, M., Guo, Y. and Hu, H., 2021. Numerical Study on Airfoil Aerodynamics in Proximity to Wavy Water Surface for Various Amplitudes. *Applied Sciences*, *11*(9), p.4215.

Mohamad, B., Karoly, J. and Zelentsov, A., 2020. CFD modelling of formula student car intake system. *Facta Universitatis, Series: Mechanical Engineering*, *18*(1), pp.153-163.

Rarata, Z., 2022. The effect of wavy surface on boundary layer instabilities of an airfoil. *Aircraft Engineering and Aerospace Technology*, *94*(2), pp.125-139.

Singh, S.K. and Narayanan, S., 2023. On the reductions of airfoil–turbulence noise by curved wavy serrations. *Physics of Fluids*, *35*(7).

Smith, T.A. and Klettner, C.A., 2022. Airfoil trailing-edge noise and drag reduction at a moderate Reynolds number using wavy geometries. *Physics of Fluids*, *34*(11).

Sun, X., Zhang, C., Shen, C., Cheng, W., Cui, Z., Wu, Z., Chen, Z. and Zhao, L., 2024. Reduction of interaction noise using grooved cylinder and wavy leading edge airfoil. *Journal of Fluids and Structures*, *125*, p.104082.

Wang, D., Cai, C., Zha, R., Peng, C., Feng, X., Liang, P., Meng, K., Kou, J., Maeda, T. and Li, Q.A., 2024. Impact of Leading-Edge Tubercles on Airfoil Aerodynamic Performance and Flow Patterns at Different Reynolds Numbers. *Energies*, *17*(21), p.5518.

Wang, Y., Ma, B., Li, Y., Jiang, S., Wen, B. and Chen, X., 2022, November. The influence of the wavy leading edge on the flow over an airfoil. In *2022 International Conference on Sensing, Measurement & Data Analytics in the era of Artificial Intelligence (ICSMD)* (pp. 1-6). IEEE.

Williamson, Gregory. "Experimental wind tunnel study of airfoils with large flap deflections at low Reynolds numbers." PhD diss., University of Illinois at Urbana-Champaign, 2012.

Wu, L. and Liu, X., 2021. Dynamic stall characteristics of the bionic airfoil with different waviness ratios. *Applied Sciences*, *11*(21), p.9943.

Xin, Z., Cai, Z., Ren, Y. and Liu, H., 2022. Comparative analysis of the self-propelled locomotion of a pitching airfoil near the flat and wavy ground. *Biomimetics*, *7*(4), p.239.

Xing, Y., Chen, W., Wang, X., Tong, F. and Qiao, W., 2023a. Effect of wavy leading edges on airfoil trailing-edge bluntness noise. *Aerospace*, *10*(4), p.353.

Xing, Y., Wang, X., Chen, W., Tong, F. and Qiao, W., 2023b. Experimental Study on Wind Turbine Airfoil Trailing Edge Noise Reduction Using Wavy Leading Edges. *Energies*, *16*(16), p.5865.

Yi-Nan, Z., Hui-Jing, C. and Ming-Ming, Z., 2021. A calculation method for modeling the flow characteristics of the wind turbine airfoil with leading-edge protuberances. *Journal of Wind Engineering and Industrial Aerodynamics*, *212*, p.104613.

Yu, F.Y., Wan, Z.H., Hu, Y.S., Sun, D.J. and Lu, X.Y., 2024. Analysis of wavy leading-edge noise reduction and source mechanism in rod-airfoil interactions. *Physics of Fluids*, *36*(4).

Zhang, Y.N., Zhang, M.M., Cai, C. and Xu, J.Z., 2020. Aerodynamic load control on a dynamically pitching wind turbine airfoil using leading-edge protuberance method. *Acta Mechanica Sinica*, *36*, pp.275-289.

Zhao, X., Cheng, H. and Ji, B., 2022. LES investigation of the cavitating hydrofoils with various wavy leading edges. *Ocean Engineering*, *243*, p.110331.

Zverkov, I.D. and Kryukov, A.V., 2021. Impact onto the boundary layer on the airfoil of a small-scale aircraft system with the use of a wavy surface. Problems and Prospects. *Journal of Applied Mechanics and Technical Physics*, *62*, pp.503-518.